\PassOptionsToPackage{usenames,rgb}{xcolor}
\documentclass[a4paper,12pt]{article}
\pdfoutput=1 
\usepackage[T1]{fontenc}
\usepackage{geometry}
\usepackage{amsfonts}
\usepackage{ae,aecompl}
\usepackage{amsmath}
\usepackage{amssymb}
\usepackage{amsthm}
\usepackage[pdftex]{graphicx}
\usepackage{multicol,multirow}       
\usepackage[small]{caption}   
\usepackage{enumerate} 
\usepackage{url}
\usepackage{mathdots}
\usepackage{xcolor}
\usepackage{tikz}

\newcommand{\ie}[0]{\textit{i.e.}}

\DeclareGraphicsExtensions{{.jpeg},{.jpg},{.png},{.pdf}}
\graphicspath{{./images/},%
{../},%
}

\title{Tomographic X-ray data of carved cheese}
\author{T.A.~Bubba$^1$, M.~Juvonen$^1$, J.~Lehtonen$^1$, M.~M{\"{a}}rz$^2$,  \\
A.~Meaney$^3$, Z.~Purisha$^1$, and 
S.~Siltanen$^1$\footnote{\textit{Email addresses}: tatiana.bubba@helsinki.fi (T.A.~Bubba), markus.juvonen@helsinki.fi (M.~Juvonen), jonatan.lehtonen@helsinki.fi (J.~Lehtonen), maerz@math.tu-berlin.de (M.~M{\"{a}}rz), alexander.meaney@helsinki.fi (A.~Meaney), zenith.purisha@helsinki.fi (Z.~Purisha), samuli.siltanen@helsinki.fi (S.~Siltanen)}}
\date{\small $^1$Department of Mathematics and Statistics, University of Helsinki, Finland\\%
    $^2$Institut f{\"{u}}r Mathematik, TU Berlin, Germany\\%
    $^3$Department of Physics, University of Helsinki, Finland} 

\begin{document}

\maketitle

\abstract{\noindent This is the documentation of the tomographic X-ray data of a carved cheese slice. Data are available at {\url{www.fips.fi/dataset.php}}, and can be freely used for scientific purposes with appropriate references to them, and to this document in \url{http://arxiv.org/}. The data set consists of (1) the X-ray sinogram of a single 2D slice of the cheese slice with three different resolutions and (2) the corresponding measurement matrices modeling the linear operation of the X-ray transform. Each of these sinograms was obtained from a measured 360-projection fan-beam sinogram by down-sampling and taking logarithms. The original (measured) sinogram is also provided in its original form and resolution. }

\section{Introduction}

The main idea behind the project is to create real CT measurement data for testing sparse and limited data tomography algorithms. A sufficiently thin slice of Edam cheese   
has been carved with CT letters to recover different attenuation coefficients (calcium-containing organic 
compounds for cheese and air for the carved letters). In particular, the shape of the carved letters
makes the target rather challenging for typical sparse and limited angle tomography applications.

A video report of the data collection session is available at 
\url{https://www.youtube.com/watch?v=omMrkaCxB2E}.

\section{Contents of the data set}
\label{sec:Contents}

The data set contains the following MATLAB\footnote{MATLAB is a registered trademark of The MathWorks, Inc.} data files:
\begin{itemize}
\item  {\tt DataFull\_512x15.mat},
\item  {\tt DataLimited\_512x15.mat},
\item  {\tt DataFull\_256x15.mat},
\item  {\tt DataLimited\_256x15.mat},
\item  {\tt DataFull\_128x15.mat},
\item  {\tt DataLimited\_128x15.mat},
\item  {\tt FullSizeSinogram.mat} and
\item  {\tt {GroundTruthReconstruction.mat}}.
\end{itemize}
The first two of these files contain CT sinograms and the corresponding measurement matrices with the 
same resolution but different spanning for the angle of view: the data in files {\tt DataFull\_512x15.mat} 
lead to reconstructions with resolutions $512 \times 512$ with $15$ directions spanning the full $360$ 
degree circle; the data in files {\tt DataLimited\_512x15.mat} lead to a reconstructions with resolutions 
$512 \times 512$ with $15$ directions spanning a $90$ degree angle of view. 
Similarly, the data in files {\tt DataFull\_256x15.mat} and {\tt DataLimited\_256x15.mat} ({\tt DataFull\_128x15.mat} and {\tt DataLimited\_128x15.mat}, respectively) lead to 
reconstructions with resolutions $256 \times 256$ ($128 \times 128$, respectively) with $15$ directions spanning the full $360$ degree circle and a $90$ degree angle of view, respectively.
The data file named {\tt FullSizeSinogram.mat} includes the original (measured) sinograms of $360$  
projections, and {\tt GroundTruthReconstruction.mat} contains a high-resolution FBP reconstruction computed from the $360$-projection sinogram. Detailed contents of each data can be found below.

\bigskip\noindent
{\tt DataFull\_512x15.mat} contains the following variables:
\begin{enumerate}
\item Sparse matrix {\tt A} of size $14835 \times 262144$; measurement matrix.
\item Matrix {\tt m} of size $989 \times 15$; sinogram ($15$ projections spanning the full $360$ 
degree circle).
\item Scalar {\tt normA}; norm of the matrix {\tt A}.
\end{enumerate}

\bigskip\noindent
{\tt DataLimited\_512x15.mat} contains the following variables:
\begin{enumerate}
\item Sparse matrix {\tt A} of size $14835 \times 262144$; measurement matrix.
\item Matrix {\tt m} of size $989 \times 15$; sinogram ($15$ projections spanning the range 
$1^{\circ}$-- $90^{\circ}$, \ie{}, a limited angle of view).
\item Scalar {\tt normA}; norm of the matrix {\tt A}.
\end{enumerate}

\bigskip\noindent
{\tt DataFull\_256x15.mat} contains the following variables:
\begin{enumerate}
\item Sparse matrix {\tt A} of size $14835 \times 65536$; measurement matrix.
\item Matrix {\tt m} of size $989 \times 15$; sinogram ($15$ projections spanning the full $360$ degree 
circle).
\item Scalar {\tt normA}; norm of the matrix {\tt A}.
\end{enumerate}

\bigskip\noindent
{\tt DataLimited\_256x15.mat} contains the following variables:
\begin{enumerate}
\item Sparse matrix {\tt A} of size $14835 \times 65536$; measurement matrix.
\item Matrix {\tt m} of size $989 \times 15$; sinogram ($15$ projections spanning the range 
$1^{\circ}$-- $90^{\circ}$, \ie{}, a limited angle of view).
\item Scalar {\tt normA}; norm of the matrix {\tt A}.
\end{enumerate}

\bigskip\noindent
{\tt DataFull\_128x15.mat} contains the following variables:
\begin{enumerate}
\item Sparse matrix {\tt A} of size $14835 \times 16384$; measurement matrix.
\item Matrix {\tt m} of size $989 \times 15$; sinogram ($15$ projections spanning the full $360$ degree 
circle).
\item Scalar {\tt normA}; norm of the matrix {\tt A}.
\end{enumerate}

\bigskip\noindent
{\tt DataLimited\_128x15.mat} contains the following variables:
\begin{enumerate}
\item Sparse matrix {\tt A} of size $14835 \times 16384$; measurement matrix.
\item Matrix {\tt m} of size $989 \times 15$; sinogram ($15$ projections spanning the range 
$1^{\circ}$-- $90^{\circ}$, \ie{}, a limited angle of view).
\item Scalar {\tt normA}; norm of the matrix {\tt A}.
\end{enumerate}

\bigskip\noindent
{\tt FullSizeSinograms.mat} contains the following variables:
\begin{enumerate}
\item Matrix {\tt sinogram15FullView} of size $2240 \times 15$; original (measured) sinogram of $15$ 
projections spanning the full $360$ degree circle.
\item Matrix {\tt sinogram15LimitedView} of size $2240 \times 15$; original (measured) sinogram of 
$15$ projections spanning the range $1^{\circ}$-- $90^{\circ}$ (limited angle of view).
\item Matrix {\tt sinogram45FullView} of size $2240 \times 45$; original (measured) sinogram of $45$ 
projections spanning the full $360$ degree circle.
\item Matrix {\tt sinogram45LimitedView} of size $2240 \times 45$; original (measured) sinogram of 
$45$ projections spanning the range $1^{\circ}$-- $90^{\circ}$ (limited angle of view).
\item Matrix {\tt sinogram180FullView} of size $2240 \times 180$; original (measured) sinogram of $180$ 
projections spanning the full $360$ degree circle.
\item Matrix {\tt sinogram360} of size $2240 \times 360$; original (measured) sinogram of $360$ projections.
\end{enumerate}

\noindent
{\tt {GroundTruthReconstruction.mat}} contains the following variables:
\begin{enumerate}
\item Matrix {\tt FBP360} of size $2000 \times 2000$; a high-resolution filtered back-projection 
reconstruction computed from the larger sinogram of 360 projections of the carved cheese 
(``ground truth''). See Figure~\ref{fig:FBP360}.
\end{enumerate}
Also, we provide the user with the MATLAB routine producing the filtered back-projection reconstruction (script file {\tt FBPgroundtruthRec.m}). 

\bigskip

\noindent In addition, at \url{www.fips.fi/dataset.php} are available data set with 45 projections (spanning both the full 360 degree circle and the range $1^{\circ}$--$90^{\circ}$) and 180 projections (spanning the full 360 degree circle), for 
the same three resolutions $512 \times 512$, $256 \times 256$ and $128 \times 128$. Similarly to the 15 projections case, the data set names are {\tt DataFull\_512x45.mat}, {\tt DataFull\_256x45.mat} and  
{\tt DataFull\_128x45.mat} for the 45 projections case spanning the full 360 degree circle, 
{\tt DataLimited\_512x45.mat}, {\tt DataLimited\_256x45.mat} and {\tt DataLimited\_128x45.mat} for the 45 limited data case (range $1^{\circ}$--$90^{\circ}$), and {\tt DataFull\_512x180.mat}, {\tt DataFull\_256x180.mat} and  
{\tt DataFull\_128x180.mat} for the 180 projections case spanning the full 360 degree circle. All these data sets 
contains the measurement matrix {\tt A} along with its norm {\tt normA} and the sinogram {\tt m}. 

\bigskip

\noindent \textbf{Remark.}
{\itshape The resolutions of the above datasets are designed specifically so that the total variation regularization parameter choice rule published in~\cite{Niinimaki2016} can be applied easily. Also, the reason to include the norms of the matrices is the following. Some reconstruction methods require that the norm of the system matrix in equation $Ax=m$ is (at most) one. This can be easily enforced like this in MATLAB:
\begin{verbatim}
A = A/normA;
m = m/normA;
\end{verbatim}
After these lines of code the equation is equivalent to the original but the norm of the system matrix is 
one.}

Details on the X-ray measurements are described in Section \ref{sec:Measurements} below.
The model for the CT problem is
\begin{equation}\label{eqn:Axm}
 {\tt A*x=m(:)},
\end{equation}
where {\tt m(:)} denotes the standard vector form of matrix {\tt m} in MATLAB and {\tt x} is the reconstruction in vector form. In other words, the reconstruction task is to find a vector {\tt x} that (approximately) satisfies \eqref{eqn:Axm} and possibly also meets some additional regularization requirements.
A demonstration of the use of the data is presented in Section~\ref{sec:Demo}.

\section{X-ray measurements}
\label{sec:Measurements}

\begin{figure}[t]
\centering
\includegraphics[width=350pt]{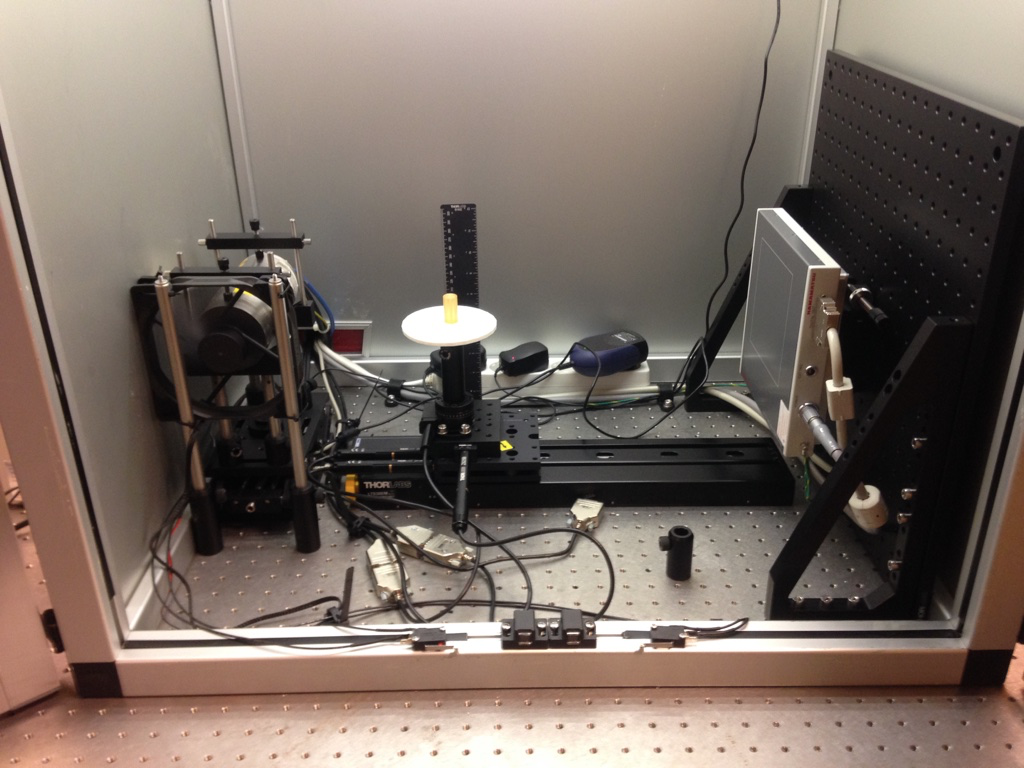}
\caption{The custom-made measurement device at University of Helsinki.}
\label{fig:Lab}
\end{figure}

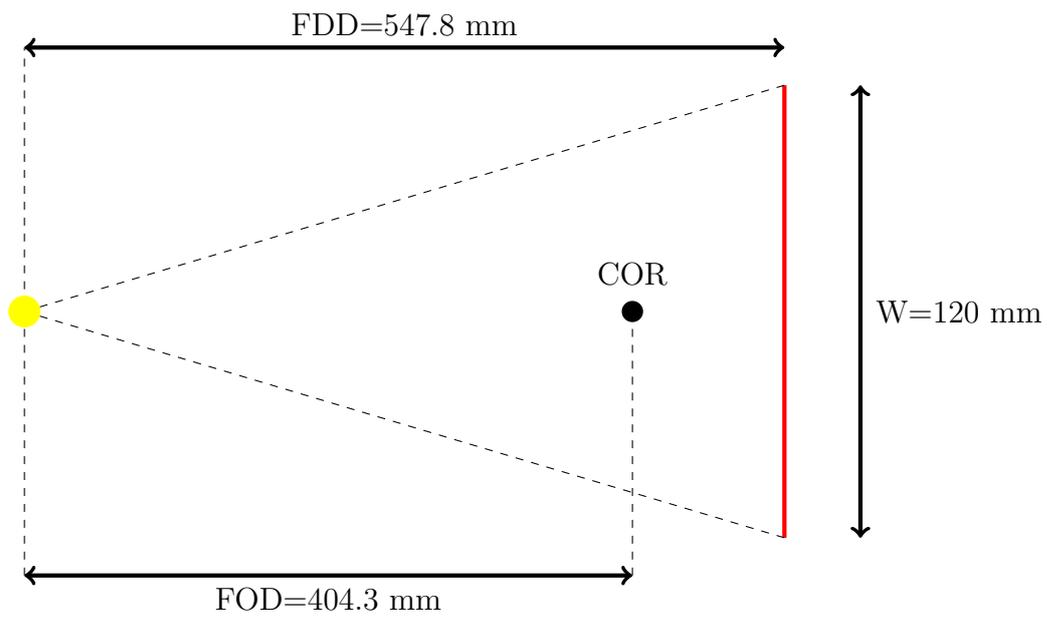
\begin{figure}
\begin{tikzpicture}[scale=2.0]
\draw[ultra thick,red] (2.5,-1.5) -- (2.5,1.5);  
\draw[<->, ultra thick] (-2.5,1.75) -- (2.5,1.75) node[above,midway]{FDD=547.8 mm}; 
\draw[<->, ultra thick] (-2.5,-1.75) -- (1.5,-1.75) node[below,midway]{FOD=404.3 mm}; 
\draw[<->, ultra thick] (3.0,-1.5) -- (3.0,1.5); 
\draw[dashed] (-2.5,-1.75) -- (-2.5,1.75); 
\draw[dashed] (1.5,-1.75) -- (1.5,0.0); 
\draw[dashed] (-2.5,0.0) -- (2.5,-1.5);
\draw[dashed] (-2.5,0.0) -- (2.5,1.5); 
\draw (3.65,0.0) node{W=120 mm}; 
\fill[thick] (1.5,0.0) circle (2pt);
\draw (1.5,0.25) node{COR};
\fill[thick, yellow] (-2.5,0.0) circle (3pt);
\end{tikzpicture}
\bigskip
\caption{Geometry of the measurement setup. Here FOD and FDD denote the focus-to-object distance and the focus-to-detector distance, respectively; the black dot COR is the center-of-rotation. The width of the detector (\ie{}, the red thick line) is denoted by W. The yellow dot is the X-ray source. To increase clarity, the $x$-axis and $y$-axis in this image are not in scale.}\label{k1}
\end{figure}

The data  in the sinograms are X-ray tomographic (CT) data of a 2D cross-section of the carved cheese measured with a custom-built $\mu$CT device shown in Figure~\ref{fig:Lab}. 
\begin{itemize}
\item The X-ray tube is a model XTF5011 manufactured by Oxford Instruments. This
model is no longer sold by Oxford Instruments, although they have newer, similar models available. The tube uses a molybdenum (Z = 42) target. 
\item The sample manipulator consists of a motorized rotation stage (Thorlabs CR1/M-Z7) mounted onto a motorized horizontal translation stage (Thorlabs LTS300/M), both of which are computer-controlled. The horizontal translation stage allows control of the geometric magnification parameter.
\item The flat panel detector is a Hamamatsu Photonics C7942CA-22. The active area of the flat panel detector is 120 mm $\times$ 120 mm. It consists of a $2400 \times 2400$ array of 50 $\mu$m pixels. According to the manufacturer the number of active pixels is  $2344 \times 2240$, which yields an active width of 112 mm. However, the image files actually generated by the camera were $2368 \times 2240$ pixels in size. 
\end{itemize}
The measurement setup was designed and assembled in 2015 by Alexander Meaney as a MSc thesis 
project~\cite{Meaney2015} and recently upgraded (1-3/2017). 
The setup is illustrated in Figure~\ref{fig:Lab} and the measurement geometry is shown in 
Figure~\ref{k1}. 

\begin{figure}[t]
\begin{picture}(390,180)
\linethickness{0.2mm}
\put(0,0){\includegraphics[height=205pt, width=300pt]{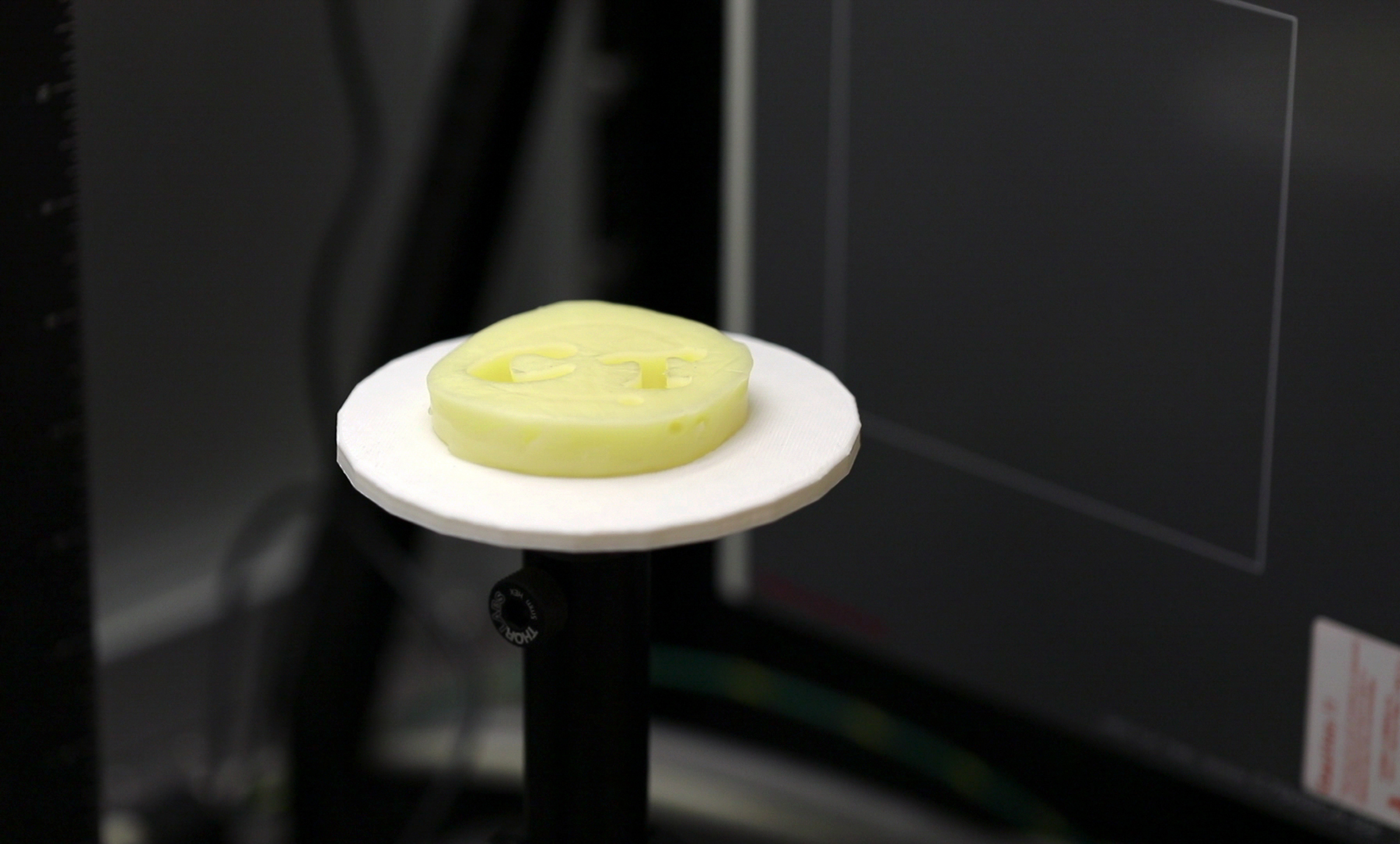}}
\put(310,0){\includegraphics[height=110pt]{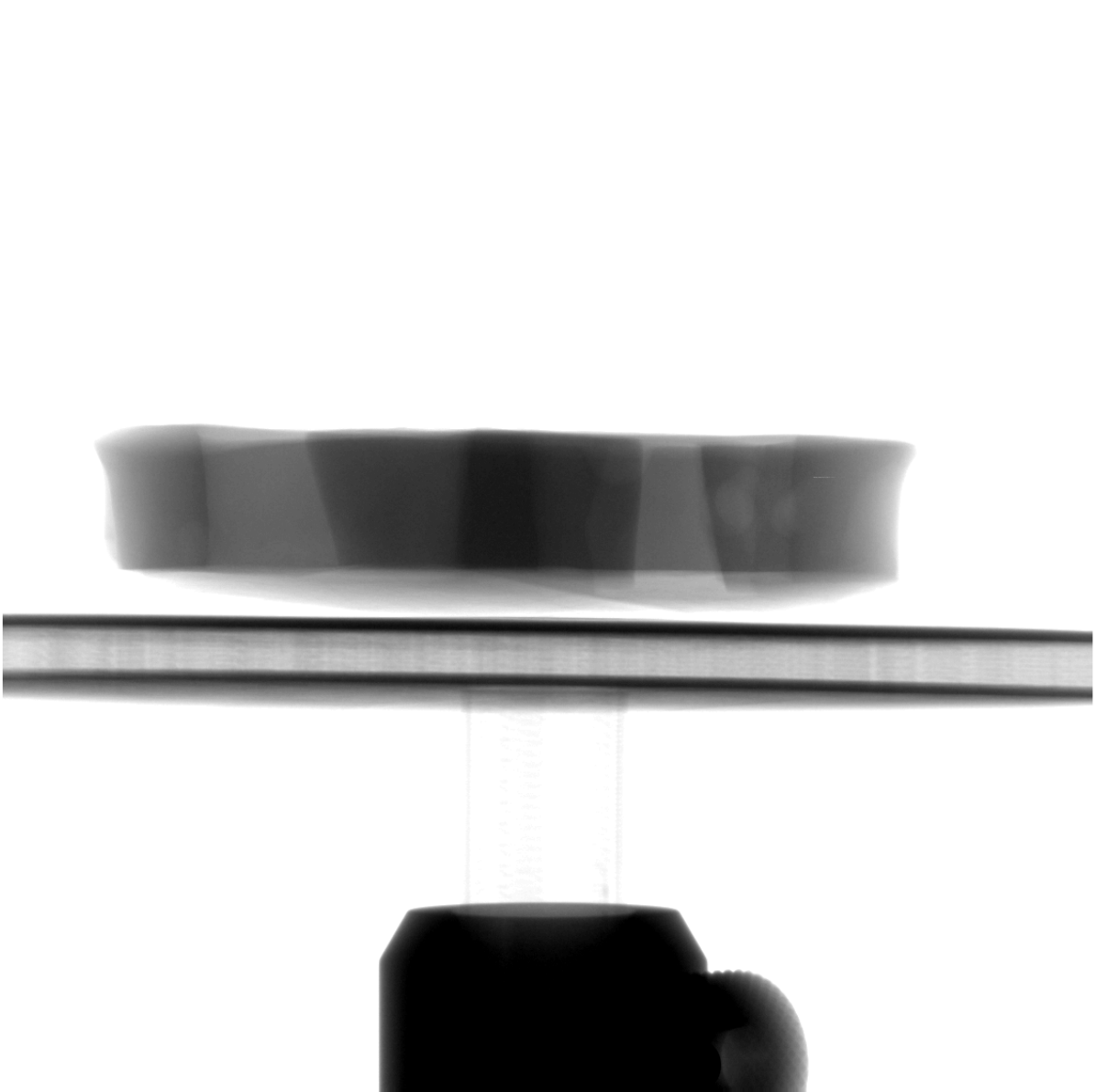}}
\put(310,0){\line(0,1){100}}
\put(310,100){\line(1,0){110}}
\put(420,100){\line(0,-1){100}}
\put(420,0){\line(-1,0){110}}
\put(310,105){\includegraphics[height=110pt]{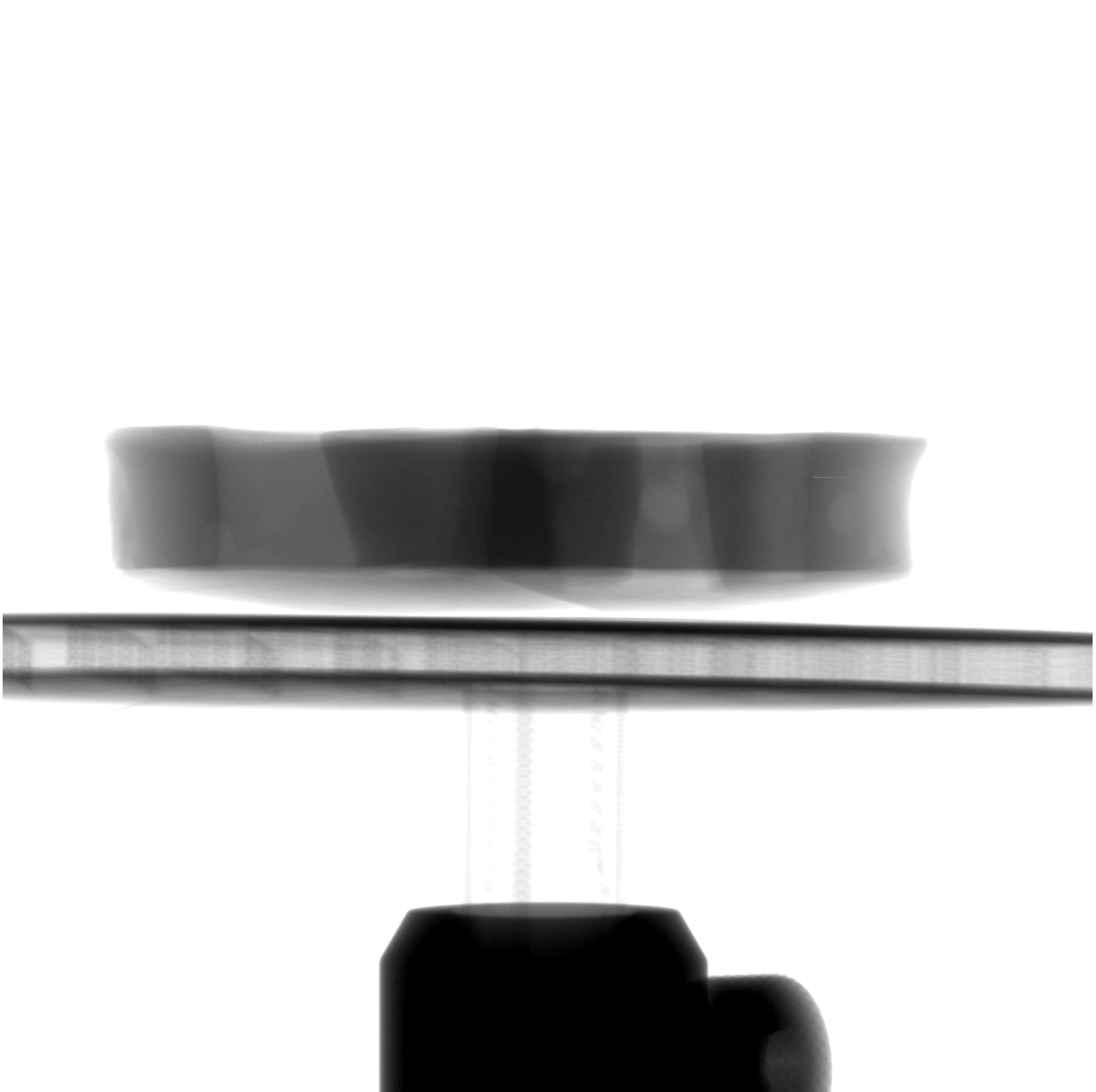}}
\put(310,105){\line(0,1){100}}
\put(310,205){\line(1,0){110}}
\put(420,205){\line(0,-1){100}}
\put(420,105){\line(-1,0){110}}
\end{picture}
\caption{\emph{Left}: Experimental setup used for collecting tomographic X-ray data. The detector plane is behind the cheese target with the active area indicated by a white square. The target is attached to a computer-controlled rotator platform. \emph{Right}: Two examples of the resulting 2D projection images. The fan-beam data in the sinograms consist of the (down-sampled) central rows of the 2D projection images.}
\label{fig:SetupAndProjections}
\end{figure}
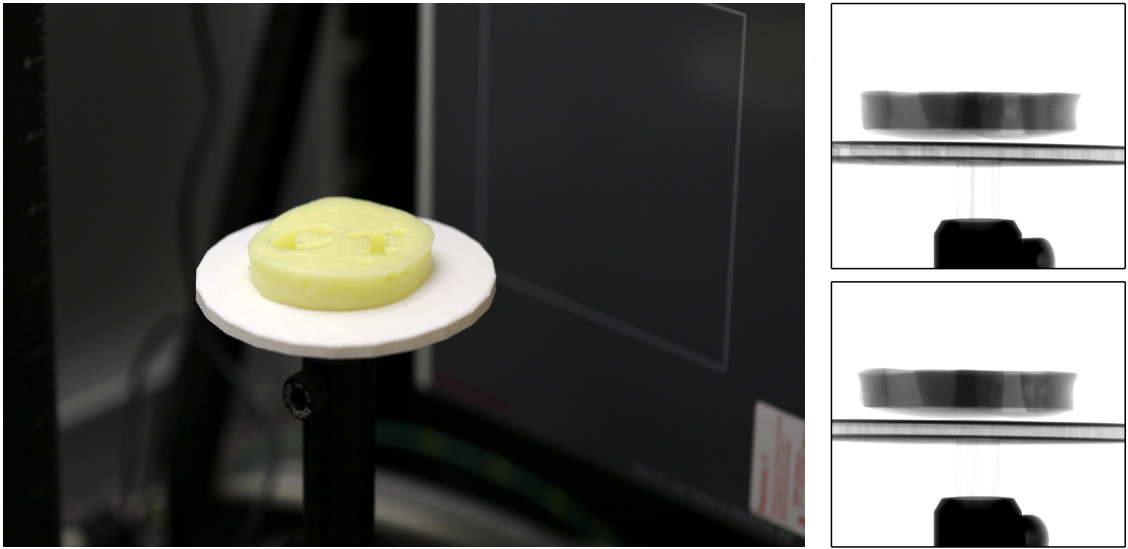

A  set of $360$ cone-beam projections with resolution $2368 \times 2240$ and angular step of one 
($1$) degree was measured. The exposure time was $1000$ ms (\ie{}, one second). The X-ray tube acceleration voltage was $40$ kV and tube current $1$\,mA. See Figure~\ref{fig:SetupAndProjections} 
for two examples of the resulting projection images. 

From the 2D projection images the middle rows corresponding to the central horizontal cross-section of the carved cheese were taken to form a fan-beam sinogram of resolution $2240 \times 15$ 
(variables {\tt sinogram15FullView} and {\tt sinogram15LimitedView} in file {\tt FullSizeSinograms.mat}). 
These sinograms were further down-sampled by binning and taken logarithms of to obtain the 
sinograms {\tt m} in all the files specified in Section~\ref{sec:Contents}.

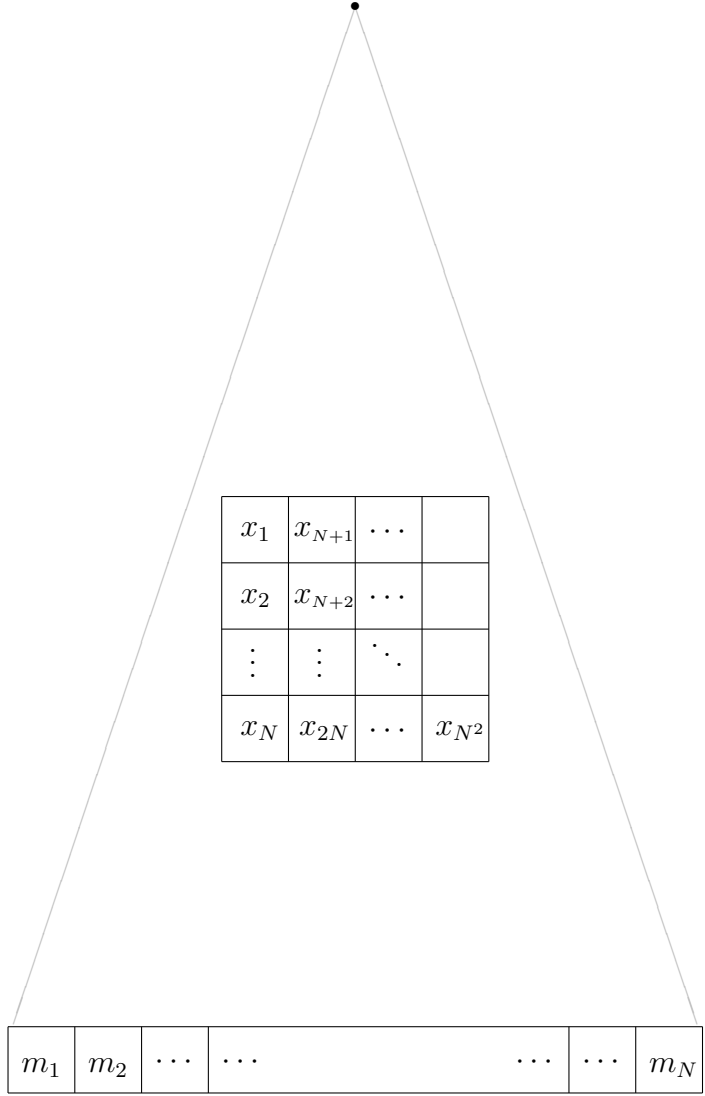
\begin{figure}
\begin{picture}(390,390)

\put(195,415){\color{lightgray}\line(1,-3){128}}
\put(195,415){\color{lightgray}\line(-1,-3){128}}

\put(195,415){\circle*{3}}

\put(145,130){\line(1,0){100}}
\put(145,155){\line(1,0){100}}
\put(145,180){\line(1,0){100}}
\put(145,205){\line(1,0){100}}
\put(145,230){\line(1,0){100}}
\put(145,130){\line(0,1){100}}
\put(170,130){\line(0,1){100}}
\put(195,130){\line(0,1){100}}
\put(220,130){\line(0,1){100}}
\put(245,130){\line(0,1){100}}
\put(152,215){$x_1$}
\put(152,190){$x_2$}
\put(155,162){$\vdots$}
\put(152,140){$x_N$}
\put(172,215){$x_{\scriptscriptstyle N+1}$}
\put(172,190){$x_{\scriptscriptstyle N+2}$}
\put(180,162){$\vdots$}
\put(174,140){$x_{2N}$}
\put(200,214){$\cdots$}
\put(200,189){$\cdots$}
\put(200,165){$\ddots$}
\put(200,139){$\cdots$}
\put(225,140){$x_{N^2}$}

\put(65,5){\line(1,0){260}}
\put(65,30){\line(1,0){260}}
\put(65,5){\line(0,1){25}}
\put(90,5){\line(0,1){25}}
\put(115,5){\line(0,1){25}}
\put(140,5){\line(0,1){25}}
\put(275,5){\line(0,1){25}}
\put(300,5){\line(0,1){25}}
\put(325,5){\line(0,1){25}}
\put(70,13){$m_1$}
\put(95,13){$m_2$}
\put(120,14){$\cdots$}
\put(145,14){$\cdots$}
\put(255,14){$\cdots$}
\put(280,14){$\cdots$}
\put(305,13){$m_N$}
\end{picture}
\caption{The organization of the pixels in the sinograms {\tt m}\,=\,$[m_1,m_2,\ldots,m_{15K}]^T$ and 
the reconstructions {\tt x}\,=\,$[x_1,x_2,\ldots,x_{N^2}]^T$, with $N=128$, $N=256$ or $N=512$. 
The picture shows the organization for the first projection. After that, in the full angular view case, the target takes a $24$ degree steps counter-clockwise (or, equivalently, the source and detector take a $24$ degree steps clockwise) and the following columns of {\tt m} are determined in a similar way; in the 
limited angular view case, the target takes a $6$ degree steps counter-clockwise (or, equivalently, the source and detector take a $6$ degree steps clockwise).
}

\label{fig:pixelDemo}
\end{figure}

The organization of the pixels in the sinograms and the reconstructions is illustrated in 
Figure~\ref{fig:pixelDemo}. 

\begin{figure}
\includegraphics[width=\textwidth]{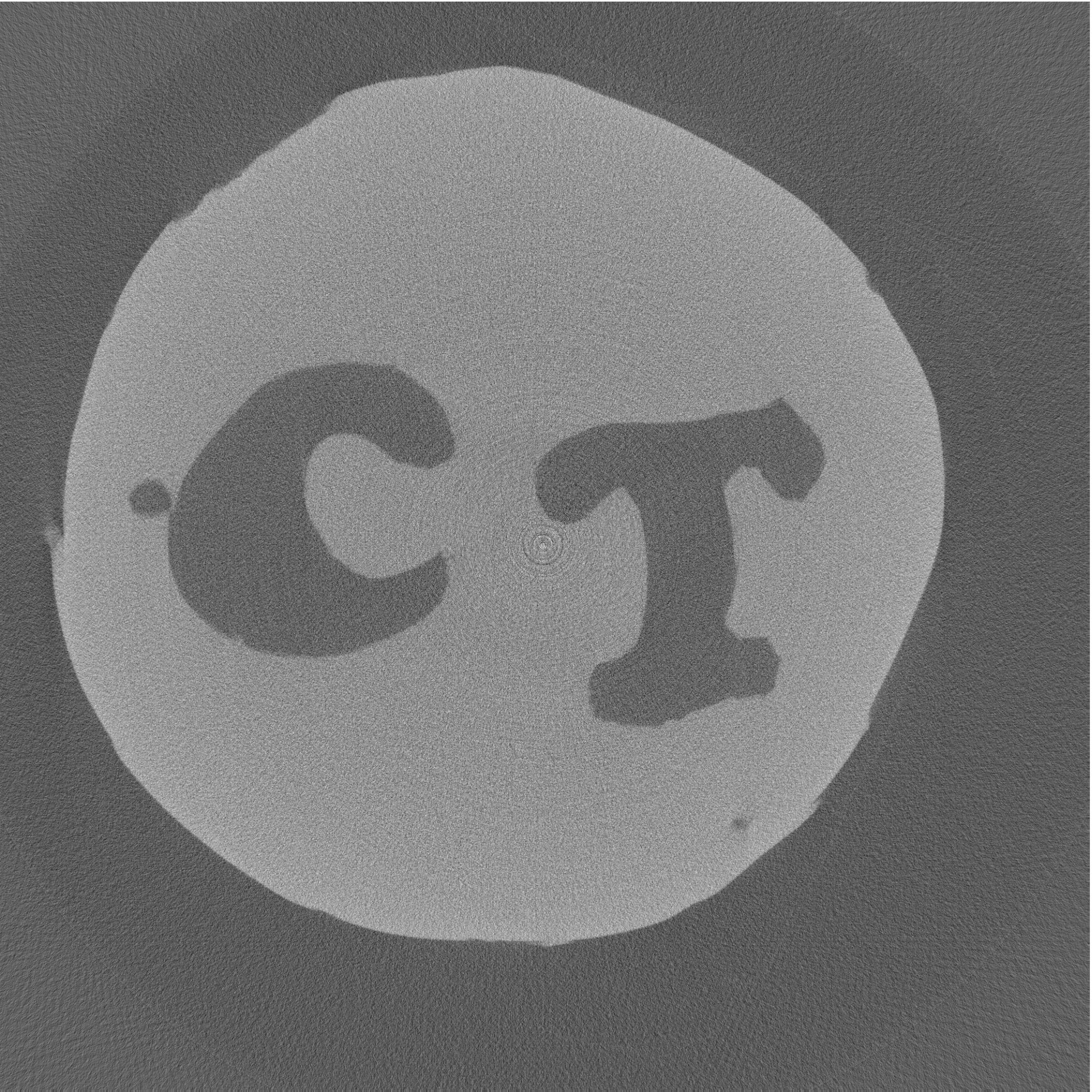}
\caption{The high-resolution filtered back-projection reconstruction {\tt FBP360} of the carved cheese computed from 360 projections.}\label{fig:FBP360}
\end{figure}

In addition, a larger set of $360$ projections of the same carved cheese using the same imaging setup and measurement geometry, but with a finer angular step of one $(1)$ degrees, was measured 
(variable {\tt sinogram360} in file {\tt FullSizeSinograms.mat}). The high-resolution ground truth reconstruction (variable {\tt FBP360} in file {\tt GroundTruthReconstruction.mat}) was computed from this data using filtered back-projection algorithm, see Figure~\ref{fig:FBP360}.

\clearpage
\section{Example of using the data}\label{sec:Demo}

\begin{figure}
\includegraphics[width=\textwidth]{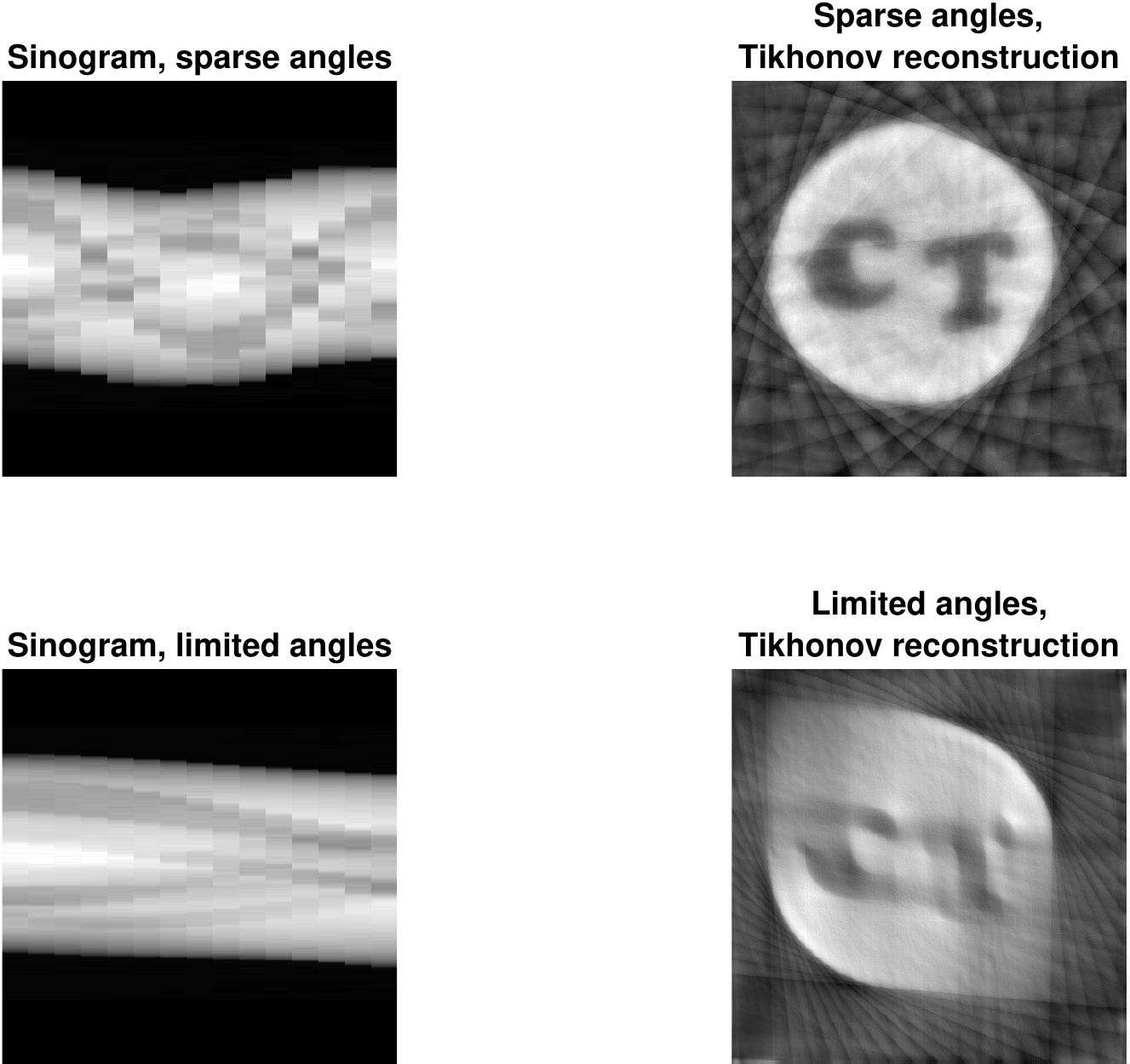}
\bigskip
\caption{First row: sinogram and corresponding Tikhonov regularized reconstruction with 15 projections spanning the full 360 degree circle. Second row: sinogram and corresponding Tikhonov regularized reconstruction with 15 projections spanning the range $1^{\circ}$-- $90^{\circ}$.}
\label{fig:Tikh}
\end{figure}

The following MATLAB code demonstrates how to use the data. The code is also provided as the separate MATLAB script file {\tt example.m} and it assumes the data files (or in this case at least the 
files {\tt DataFull\_512x15.mat} and {\tt DataLimited\_512x15.mat}) are included in the same directory with the script file. The resulting reconstructed images are reported in Figure~\ref{fig:Tikh}.

\begin{verbatim}
% Load the measurement matrix and the sinogram for the sparse angle case
load DataFull_512x15
N1    = sqrt(size(A,2));
m1    = m;

% Compute a Tikhonov regularized reconstruction using
% conjugate gradient algorithm pcg.m
alpha = 10; % regularization parameter
fun   = @(x) A.'*(A*x)+alpha*x;
b     = A.'*m(:);
x1    = pcg(fun,b);

% Load the measurement matrix and the sinogram for the limited angle case
load DataLimited_512x15
N2    = sqrt(size(A,2));
m2    = m;

% Compute a Tikhonov regularized reconstruction using
% conjugate gradient algorithm pcg.m
alpha = 10; % regularization parameter
fun   = @(x) A.'*(A*x)+alpha*x;
b     = A.'*m(:);
x2    = pcg(fun,b);

% Take a look at the sinograms and the reconstructions
figure
subplot(2,2,1)
imagesc(m1)
colormap gray
axis square
axis off
title('Sinogram, sparse angles')
subplot(2,2,3)
imagesc(m2)
colormap gray
axis square
axis off
title('Sinogram, limited angles')
subplot(2,2,2)
imagesc(reshape(x1,N1,N1))
colormap gray
axis square
axis off
title({'Sparse angles'; 'Tikhonov reconstruction,'})
subplot(2,2,4)
imagesc(reshape(x2,N2,N2))
colormap gray
axis square
axis off
title({'Limited angles'; 'Tikhonov reconstruction,'})
\end{verbatim}

\end{document}